\documentclass[manuscript]{acmart}
\AtBeginDocument{%
  \providecommand\BibTeX{{%
    \normalfont B\kern-0.5em{\scshape i\kern-0.25em b}\kern-0.8em\TeX}}}

\begin{document}

\newcommand\systemname{Twigs}

\title[Text-Based UI Mock-up Generation using Deep Learning]{Creating User Interface Mock-ups from High-Level Text Descriptions with Deep-Learning Models}

\author{Forrest Huang}
\authornote{Work done during internship at Google Research.}
\affiliation{%
\institution{University of California, Berkeley}
\city{Berkeley}
\country{USA}}
\email{forrest_huang@berkeley.edu}

\author{Gang Li}
\affiliation{
\institution{Google Research}
\city{Mountain View}
\country{USA}
}
\email{leebird@google.com}

\author{Xin Zhou}
\affiliation{
\institution{Google Research}
\city{Mountain View}
\country{USA}
}
\email{zhouxin@google.com}

\author{John F. Canny}
\affiliation{%
\institution{University of California, Berkeley}
\city{Berkeley}
\country{USA}}
\email{canny@berkeley.edu}

\author{Yang Li}
\affiliation{
\institution{Google Research}
\city{Mountain View}
\country{USA}
}
\email{liyang@google.com}




\renewcommand{\shortauthors}{F. Huang, et al.}
\newcommand\draft[1]{{\color{purple}{#1}}}
\newcommand\addfig[1]{{\color{teal}[Figure here: (#1)]}}
\newcommand\CN{{\color{blue}[Citation Needed]}}

\begin{abstract}
  The design process of user interfaces (UIs) often begins with articulating high-level design goals. Translating these high-level design goals into concrete design mock-ups, however, requires extensive effort and UI design expertise. To facilitate this process for app designers and developers, we introduce three deep-learning techniques to create low-fidelity UI mock-ups from a natural language phrase that describes the high-level design goal (e.g. ``pop up displaying an image and other options''). In particular, we contribute two retrieval-based methods and one generative method, as well as pre-processing and post-processing techniques to ensure the quality of the created UI mock-ups. We quantitatively and qualitatively compare and contrast each method's ability in suggesting coherent, diverse and relevant UI design mock-ups. We further evaluate these methods with 15 professional UI designers and practitioners to understand each method's advantages and disadvantages. The designers responded positively to the potential of these methods for assisting the design process.
 \end{abstract}

\ccsdesc[500]{Human-centered computing~Human computer interaction (HCI)}

\keywords{machine learning, text, user experience design, transformer, interaction design, deep learning}

\begin{teaserfigure}
    \includegraphics[width=\textwidth]{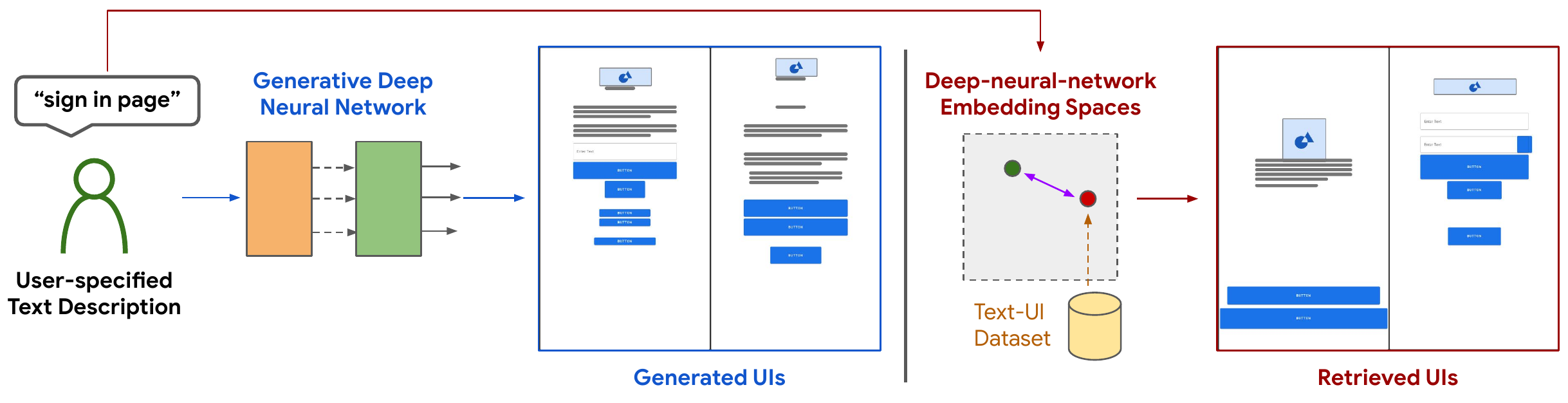}
    \caption{We introduce three novel methods for creating UI mock-ups from high-level text descriptions with deep neural networks.}
    \Description{This figure outlines the workflow of our proposed methods of creating UI mock-ups from text descriptions. On the left of the screen, there is a user icon providing the text description of ``sign in page''. This text description is passed through two separate methods. One of them is a generative deep neural network and then generate user interface with various icons and text fields similar to a mobile login screen from scratch. Another method embeds this text using a deep-neural-network into an embedding space and retrieves similar UIs from a paired text-UI dataset.}
\end{teaserfigure}

\maketitle

\section{Introduction}
Designing user interfaces is a complex process that requires significant domain-specific expertise gathered from years of education and experience. One critical step of this process is to produce low-fidelity mock-up prototypes from high-level specifications of requirements \cite{doet}, and this step often involves professional designers dedicating extensive time and effort. Many small-scale app developers, however, might not have the resources and/or expertise for such a design task, leading to lower quality end-products. While producing these mock-ups from scratch might be difficult, specifying high-level requirements for these UIs in natural language requires less design expertise. These specifications might even already exist in these developers' current development processes. 

As such, we believe computational systems that can create concrete UI designs mock-ups guided by text descriptions can greatly benefit these UI design practitioners. Moreover, these computational systems can be helpful in communicating and discussing designs with clients or other non-experts, as the generated design artifacts can be useful for them to ground their discussions on. Producing diverse sets of UIs relevant to specifications still takes significant effort and time even for professional designers, and such system can help reduce professional designers' workload, allowing them to make more important decisions in the design process. 

The first step towards such a system is developing methods that can robustly, coherently and diversely generate plausible UI design mock-ups from brief text descriptions about the desired UIs. While there have been numerous attempts at combining text and visual design in the HCI research community, many of them focus on either managing user-curated, inspirational design artifacts linked to text concepts \cite{metamap}, or generating designs using linguistic-parser-based non-neural methods \cite{crosspower}. The former type of techniques would suggest designs limited to narrow sets of inspirations already considered by designers, while the latter methods often produce layouts related to rigid linguistic structures and might not handle high-level, flexible concepts well. Researchers also attempted to crowd-source designs using text descriptions \cite{apparition}, but relying on crowd-workers might affect the quality of final designs in real-world contexts.

Recent advancements in the deep-learning community contributed text-to-image retrieval and generation models that improved the ability of machines to generate and retrieve high-quality, plausible and realistic visual content based on text descriptions. Large datasets of pairs of UIs and captions have also been recently introduced~\cite{screen2words}. With these large-scale datasets and improved deep-learning models, we introduce three deep-learning models\footnote{All implementations and weights of the models presented in this paper will be open-sourced on GitHub.} which are all first of their classes to be able to create UI mock-ups based on a wide-range of natural language descriptions through either retrieval or generation:

\begin{itemize}

\item we introduce \emph{UI Generator}, the first deep generative model that is able to generate UI mock-ups from scratch with only a high-level text description about the desired UI, and a set of post-processing techniques to filter and present quality UI designs to users.
\item we introduce \emph{Multi-modal Retriever}, the first deep-learning model that learns cross-modality correspondence and latent representation to retrieve design examples from a large UI corpus using a high-level text description about the desired UIs.
\item we introduce \emph{Text-only Retriever}, the first deep-learning method that retrieves design examples from a UI corpus based on the similarity between the text description coupled to each UI in the corpus and a high-level description about the desired UI.

\end{itemize}

We quantitatively and qualitatively compare and contrast the UIs created by these methods using automatic metrics and in a user study with 15 experienced UI designers and practitioners. We found that each of the proposed methods has unique strengths and weaknesses in different design scenarios, which allows these methods to cover various needs for assistance in UI design processes. With these proposed methods, we hope to contribute the important building blocks to future systems that can facilitate UI design processes for expert UI designers and non-expert app developers. We believe both groups can make more informed design decisions when developing mobile apps using design artifacts suggested by our proposed methods, leading to them creating better end-user experiences.
\section{Related Work}
Generating and retrieving various visual content using text descriptions are important applications extensively investigated by HCI, Computer Vision and Machine Learning researchers. This section surveys a selection of prior work that either achieve such text-to-visual ability using different methodologies, or contribute important building blocks to our proposed methods.

\subsection{Multi-Modal UI Embeddings and Text-to-Image Retrieval}
Deep-learning models have recently been successful in accomplishing information retrieval tasks under the same domain, particularly for those that involve high-dimensional visual content, such as UI designs. The state-of-the-art for UI-to-UI retrieval methods is Screen2Vec~\cite{screen2vec}, which developed semantically meaningful embeddings by training deep-learning models on multi-modal data (e.g., element types, text content) of the UIs. These low-dimensional embeddings are subsequently used to represent individual UIs and can be used for nearest-neighbour-retrieval using common distance metrics, such as cosine similarity. While Screen2Vec is primarily designed for UI-to-UI retrieval, it introduced a baseline that only uses BERT embeddings~\cite{devlin-etal-2019-bert} of text content within UIs to embed them. This is closely related to one of our methods, Text-only Retriever, because it theoretically enables UI retrieval from arbitrary text. However, these baseline embeddings depend only on actual text content in the UIs which might not correspond to matching high-level design concepts of the UIs (see our findings in Section~\ref{sec:qual}). In contrast, our retrieval-based methods allow users to specify short, high-level descriptions to retrieve UIs.

Deep-learning researchers also investigated cross-modality retrieval between text and visual content. A recent research work in this area is the dual-encoder that uses contrastive learning to learn a common embedding space between images and their corresponding text descriptions using a batch softmax loss \cite{dualencoder}. These learned embeddings corresponding to artifacts in either modalities can subsequently be used to retrieve any text or image examples in the common embedding space. However, this method has not been previously applied to retrieving UIs with text. Our Multi-modal Retriever adapts this dual-encoder paradigm to be the first deep-neural-network capable of performing text-to-UI retrieval.

\subsection{Text-to-Visual-Content Synthesis}
Beyond retrieving existing visual content, the introduction of Generative Adversarial Networks (GANs) \cite{gan} and Transformer Networks (Transformers) \cite{transformer} significantly improved machines' ability in generating visual content from scratch. Significant research effort has consequently been devoted to creating deep-learning models that bridge text and visual content (primarily natural images) using these model architectures. The current state-of-the-art of GAN-based methods is XMC-GAN \cite{xmcgan} that combines semi-supervised contrastive learning and adversarial learning to produce high-fidelity images from text.

Researchers have also recently applied Transformers to generate sequences of image regions from text tokens, which is a sequence-to-sequence translation task that Transformers have tremendous success in. The state-of-the-art Transformer-based text-to-image model is DALL-E \cite{dalle} that uses a GPT-3-based architecture \cite{gpt3} to generate discrete image codes from text descriptions. Similar to DALL-E, Scones \cite{scones} also generates visual tokens from multiple turns of text descriptions, but represents output visual objects with high-level attributes similar to our proposed UI Generator's outputs.

We chose a Transformer-based method for our UI Generator as we believe UI elements are better encoded as sequences of low-dimensional vectors (element dimensions and classes). Unlike DALL-E and Scones, the UI Generator uses a full Transformer Encoder-Decoder architecture that explicitly encodes text and decodes UI elements with separate sub-networks, which will be discussed in detail in Section \ref{sect:generator}.

\subsection{Neural Layout Generation Methods}
Other than generating natural images, various neural-network architectures, including Transformers and GANs, have also been used to generate documents and UI layouts. LayoutGAN \cite{layoutgan} utilizes a differentiable renderer to generate realistic layouts with adversarial learning. Neural Design Networks \cite{neuraldesign} took an alternative approach of encoding elements and inter-element relations of layouts as graphs and subsequently using a Graph Convolutional Neural-network to generate designs with coherent inter-element relations. 

Most closely related to our work are LayoutTransformer \cite{layouttransformer} and Variational Transformer Network (VTN) \cite{vtn}. LayoutTransformer uses a Transformer decoder-only network to decode UI elements as discrete tokens. VTN builds upon LayoutTransformer and adds an encoder to build a gaussian-like latent space for better generation performance and interpolating between UIs. While these networks can both generate diverse UI elements, our proposed UI Generator's main contribution over these prior models is that it allows users to control the UI generation process with text descriptions. To the best of our knowledge, our UI Generator is the first deep-learning model that is capable of generating UIs given text conditions.

\subsection{UI Captioning Datasets and Models}
Similar to deep-learning-based techniques introduced in the previous subsections, all of our proposed methods are based on deep-learning, which require high-quality dataset(s) relevant to their target tasks. While datasets of designer-generated UIs from text prompts are not directly available in the wild, a closely related dataset screen2words \cite{screen2words} can be repurposed for our task. Screen2words investigates the task of UI captioning, which is having crowd-workers write text descriptions after seeing UI screenshots. Large number of pairs of UIs and text descriptions collected using this method are compiled into screen2words and were initially used to train a deep-neural-network to generate these text descriptions from UIs. In our proposed methods and tasks, we can similarly use this dataset either in the opposite direction where we give the model text descriptions and train it to generate UIs, or to use contrastive-learning-based methods to learn non-directional correspondences between text and UIs.

All screenshots in screen2words are curated from the Rico \cite{rico} dataset which contains app interaction traces captured from human usage. These interaction traces contain individual states of UIs during interactive sessions in the form of screenshots and UI view hierarchies. Screen2words in addition only includes a clean subset of Rico UIs in RicoSCA \cite{ricosca}. We also combined \cite{semantics} with screen2words to obtain a paired text-UI dataset with semantically meaningful elements relevant to our use-cases.

\subsection{Interactive Text-Based Layout Synthesis}
Other than deep-learning-based methods developed by Machine Learning researchers, the HCI community have also developed non-neural layout synthesis systems for various applications. One system that has the most similar ability to ours is Apparition \cite{apparition}, in which a user can describe in text and sketch at the same time while having crowd-workers construct a UI mock-up based on them. Crosspower \cite{crosspower} is a system that is able to automatically generate slide layouts for presentations based on linguistic structures of corresponding text content, such as scripts. In the broader theme of combining text descriptions with user actions, PixelTone \cite{crosspower} allows users to combine direct manipulation methods with text commands to reduce difficulty of photo editing tasks. We share the same vision as these systems that text descriptions can be a useful medium for computational systems to understand users' high-level goals beyond specific low-level interactions. We hope that our contributed methods can be used to develop more robust and powerful systems on these applications in the future.
\section{Proposed Methods}
\begin{figure}[h]
  \centering
  \includegraphics[width=0.95\linewidth]{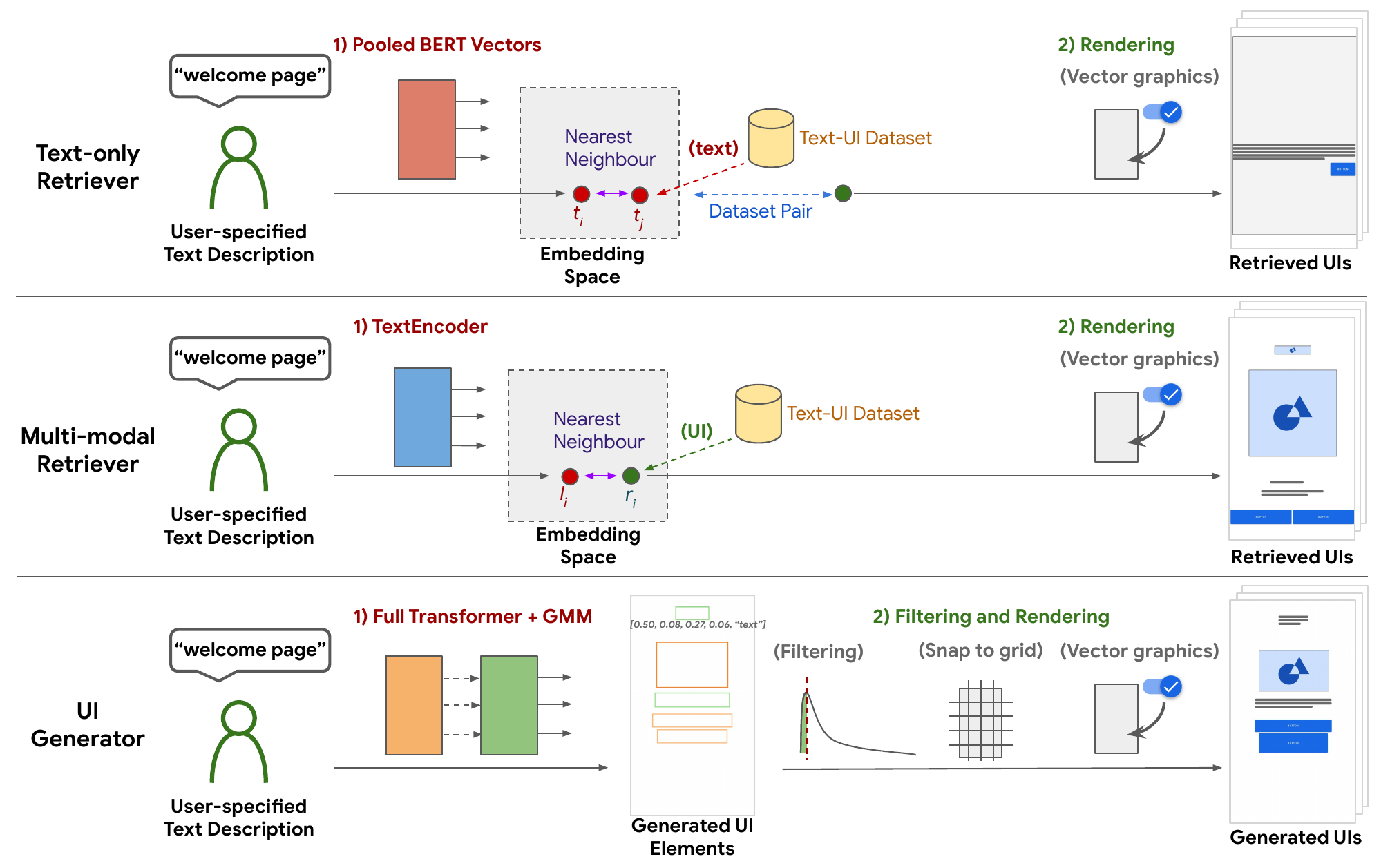}
  \caption{Side-by-side comparison of the overall workflows of each of our methods.}
  \Description{This figure was split into three section vertically and presents all three proposed methods in this paper at once. The top part of the figure presents the Text-only retriever, where a user supplies a message ``welcome page'' on the left. This message gets passed a pooled BERT vector, and then such vector is denoted as a dot in an embedding space. In this embedding space, similar text embeddings are retrieved and the corresponding UI dataset pair of this text embeddings is then used to render retrieved UIs use vector graphics. 
  The middle part of this figure presents the Multi-modal retriever, where it uses the TextEncoder the embed the same text message of 'welcome page'. However, instead of using text to retrieve nearest neighbour in the embedding space, this method directly retrieved UI embeddings as the model was able to directly embed all UIs in the dataset. This UI then gets rendered to as retrieved UIs on the left, with a UI having a large image in the middle and two buttons in the bottom.
  The bottom part of this figure presents the UI generator, where it uses a Full Transformer and Gaussian mixture model to generate attributes of UI elements on the screen. These attributes then go through filtering, snapping to grid, and finally gets rendered with vector graphics components into a similar welcome page as the rest.}
  \label{fig:overview}
\end{figure}

In this section, we describe in detail our core contribution--three methods for retrieving and generating UIs from high-level text descriptions. Figure~\ref{fig:overview} gives an overview on the commonalities and differences of all three methods. The first method is a \emph{Text-only Retriever} that retrieves the UIs in the dataset that are paired with text descriptions most similar to the input description based on the generic BERT \cite{devlin-etal-2019-bert}, a large language model. The second method is a \emph{Multi-modal Retriever}, which includes a dual-encoder \cite{dualencoder} that is capable of processing (embedding) a text description provided by a user into a common embedding space that is populated with both text and UI embeddings. We can then directly retrieve the nearest UIs in this space. Finally, the \emph{UI Generator} is a generative model that learns to synthesize UIs entirely from scratch, only from high-level text descriptions. Because the UI Generator needs to learn to compose UI mock-ups, instead of retrieving existing UIs as the other two methods, this method includes an extra post-processing step to ensure it produces high quality UIs. We will describe different processes of dataset selection and processing, model architectures, training configurations and resources required by each of these methods in the remainder of this section.


\subsection{Datasets}
While all three methods use different modalities and features for training, they all commonly use the screen2words dataset\footnote{\url{https://github.com/google-research/google-research/tree/master/screen2words}}, a large-scale dataset with more than 112,000 high-level text descriptions corresponding to detail attributes and screenshots of more than 22,000 UIs in the RicoSCA dataset. To our knowledge, it is the only large-scale dataset that contains pairs of high-level text descriptions and UI examples, which are required by our target task of text-based creation of UI mock-ups. 

We used the original dataset splits with 15712\footnote{The original screen2words training set contains 15743 UIs. However, there are 31 UIs that we were unable to process and hence we excluded them in our training set.}/2364/4310 UIs in the training/validation/test set. The UIs in each of the splits are captured from different apps. Each UI is captioned by 5 crowd-workers, producing 5 English sentences with under 10 text tokens. 

\subsection{Text-only Retriever}
High-level text descriptions of the UIs in the screen2words dataset provide brief, abstract and high-level views of the concepts and usages of the regarded UI designs. Our proposed \emph{Text-only Retriever} is a method focused only on high-level text correspondences in the dataset. It retrieves the UIs in the training set whose text description is most similar to the input text description in the embedding space. 

\subsubsection{Additional Pre-processing}
The Text-only Retriever only retrieves existing UI examples based on the similarity of text descriptions of UIs from the screen2words dataset to text descriptions of the desired UIs. We converted each text description in the training set to lower case and tokenized it according to the standard pre-processing steps for the BERT model.

\subsubsection{Model Embedding and Retrieval}
To obtain text-only embeddings, we encoded each pre-processed description using a pre-trained BERT model (uncased, hidden-layer size of 768, 12 layers) \cite{devlin-etal-2019-bert} to get a pooled embedding from the CLS token (a common practice to obtain a single fixed-length vector given a sentence). This model was trained only on BERT's original large-scale text corpuses, and was not specifically fine-tuned on domain-specific UI captions. We stored all of these embeddings as well as their corresponding UIs.

To retrieve UIs from a text description given by a user, we first pre-process and embed the description using the same BERT model and pre-processing steps applied to our training examples. We then find the nearest-neighbours of this description from all training-set descriptions in the embedding space by their Euclidean distances. Finally, we present to the user UI mock-ups created based on the specific UIs paired with these retrieved nearest-neighbour text descriptions in the dataset.

\subsection{Multi-Modal Retriever}
The screen2words dataset contains far more information than the high-level text descriptions, which can be effectively utilized by our models. As such, we developed a more advanced method that considers both text descriptions and UI details. This method embeds both text descriptions and UIs into a shared embedding space using a contrastive-learning method, and then perform retrieval across the modalities directly in this space. To our knowledge, this is the first cross-modality text-to-UI deep-learning retrieval model.

\subsubsection{Additional Pre-processing}
In contrast to the Text-only Retriever, we included a large amount of information from both the text descriptions and the paired UIs from the screen2words dataset to obtain cross-modality embeddings that encode more nuanced and detailed UI knowledge. For text descriptions, we pre-processed the individual text tokens to obtain complete sequences of BERT embeddings using the same model as the other methods. We then flattened UIs to element attribute sequences and added a start token, an end token, and a token that contains pooled BERT embedding of the text description of the UI's app to each sequence. For each element in the UI (including intermediate elements), we embed the dimensions ($x, y, w, h$), the element type (same as ones used in the screen2words dataset), and a single pooled BERT embedding of the element's text content. This gives us detailed content-based, semantic, and geometric information of each element in the UIs. We also filter out UIs that are longer than 512 tokens.

\subsubsection{Model Architecture and Training}
The Multi-modal Retriever encodes text descriptions and UIs respectively using two sub-modules $\textbf{TextEncoder}$ and $\textbf{UIEncoder}$, which is adapted from a dual-encoder \cite{dualencoder} used for text-to-image retrieval. Each of $\textbf{TextEncoder}$ and $\textbf{UIEncoder}$ is a Transformer Encoder with hidden-layer size of 64, intermediate size of 256, and 4 layers (chosen by performance on the validation set). 

Given a text description, we obtain the BERT embeddings $t_{1...k}$ for each of the $k$ text tokens from the pre-processing step. From these embeddings we can use the $\textbf{TextEncoder}$ to obtain a single, fixed-length text embedding $l = \textbf{TextEncoder}(t_{1...k})$ by taking only the output of the special start token as the `pooled vector' of the sequence. Similarly, from the pre-processing step we obtain the flattened sequence of $n$ UI element attribute vectors $u_{1...n}$. From this we can obtain a single, fixed-length UI embedding $r = \textbf{UIEncoder}(u_{1...n})$ at the start token position similar to the $\textbf{TextEncoder}$.

With pairs of corresponding $l$ and $r$, we obtain a mini-batch of $K$ pairs of embeddings $l_{1...K}$ and $r_{1...K}$. Here we formulate a loss function that would minimize the distance between matching pairs of $l$s and $r$s and maximize the distance between unmatching pairs. We minimize the following bidirectional in-batch sampled softmax loss:
$$L(\textbf{TextEncoder}, \textbf{UIEncoder}) = - \frac{1}{K} \sum_{i=1}^{K} \Big(S(l_i, r_i) - \log \sum_{j = 1, j \neq i}^{K} \exp{\big(S(l_i, r_j)\big)}\Big) - \frac{1}{K} \sum_{i=1}^{K} \Big(S(r_i, l_i) - \log \sum_{j = 1, j \neq i}^{K} \exp{\big(S(r_i, l_j)\big)}\Big)$$
given $S(l, r)$ to be the dot product between the text and UI embeddings. We use an AdamOptimizer with a constant learning rate of $0.001$ to train our model with this loss. We implemented our model using Trax\footnote{\url{https://github.com/google/trax}} and trained it on Google Cloud TPU v3 with 32 cores for ~2 days with mini-batches of 64 samples per replica.

\subsubsection{Embedding and Retrieval}
\begin{table}
  \caption{Cross-Modality Retrieval Accuracy Results}
  \label{tab:ret}
  \begin{tabular}{l|cc}
    \toprule
    &Top-1&Top-10\\
    \midrule
    \textbf{Multi-modal Retriever} (5 subsets avg.) & 23.2\% & 65.0\%\\
    SWIRE & 15.9\% & 60.9\%\\
    \midrule 
    \textbf{Multi-modal Retriever} (entire test set) & 2.80\% & 4.84\% \\
  \bottomrule
\end{tabular}
\end{table}

To retrieve UIs given a user's text description $t_{u}$, we compute $l_u = \textbf{TextEncoder}(t_{u})$ and find the nearest neighbouring UI embeddings $r$ in the training set using the dot-product similarity metric. The nearest UIs are subsequently presented to the user.

We tested this retrieval approach with embeddings of text and UIs in the test set and show retrieval accuracy\footnote{Retrieval accuracy is measured by top-k: the proportion of text descriptions that have their ground-truth corresponding UIs as one of the k nearest neighbours during retrieval.} in Table~\ref{tab:ret}. While the overall performance is relatively low, this is because the number of text and UI pairs in the entire dataset is quite high (4310 UIs $ \times$ 5 descriptions each) and similar descriptions might be paired with different UIs, which increase the difficulty of this evaluation scenario. However, when we compare our method with an established cross-modality sketch-to-UI retrieval method Swire \cite{swire}, we obtain better performance than Swire from the same number of candidates---over the average of 5 random subsets of the same size as Swire's test set (276 candidates).

\subsection{UI Generator}
\label{sect:generator}
While retrieval models can provide users with relevant and real examples in large corpuses of UIs, generative models can learn from existing designs to synthesize novel UIs that are potentially unconsidered by human designers based on unseen and flexible text descriptions. Therefore, we built a Transformer-based generative model to generate UI element attributes directly from text descriptions. 

\subsubsection{Additional Pre-processing}
While each UI example in the combined dataset of Rico and screen2words includes detailed attributes of all UI elements, not all of such information are useful for creating low-fidelity mock-ups. In contrast, requiring a model to learn to generate all UI elements and attributes drastically increases the difficulty of the task and the amount of noise in the dataset. Therefore, for our UI Generator we first filter and extract only leaf UI elements which includes most interactive components. We then flatten the set of leaf elements within each UI by sorting them based on their approximate Y coordinates. If two elements have approximately equal Y coordinates, we then sort them by their X coordinates. This creates an order where top left UI elements always exist earlier than elements in the bottom right in the flattened sequence. 

The UI element data in the original Rico dataset does not contain semantic classes of UI elements. We augment our dataset with semantic annotations published in \cite{semantics} for each UI element. This covers approximately 89.5\% of the elements in our dataset. We then reviewed the remaining 10.5\% of elements and found that 6.02\% out of the 10.5\% are separator elements that are either very narrow or very short, which we then grouped into a separate class using a heuristic. We leave the remaining elements as belonging to the unknown class but keep them in our UI element sequences. 

Other than pre-processing UI elements, we also embed text descriptions using the same pre-trained BERT model used in the other two proposed techniques. We extract one BERT embedding $t_i$ for each token $e_i$ in the entire text description sequence.

\subsubsection{Task Formulation and Model Training}

\begin{figure}[h]
  \centering
  \includegraphics[width=0.8\linewidth]{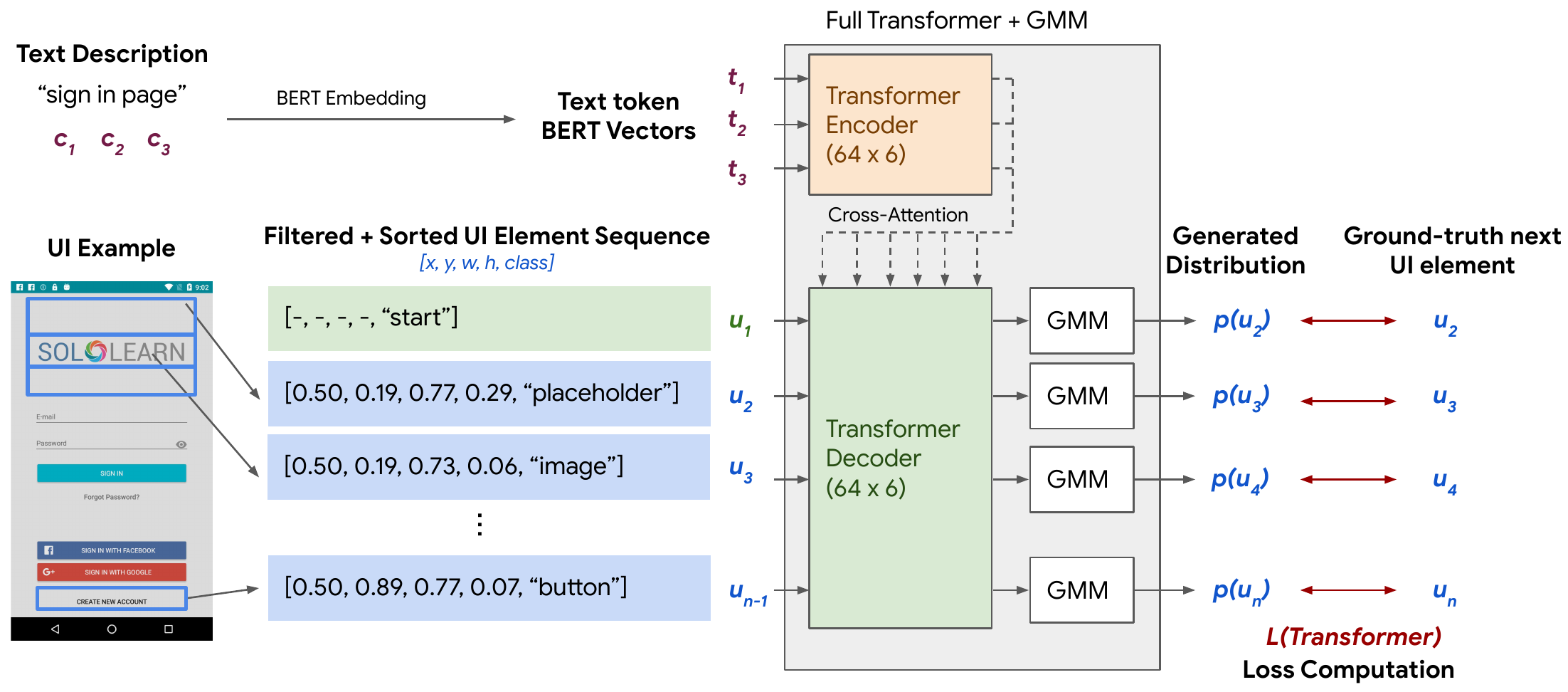}
  \caption{Model Architecture of the UI Generator}
  \Description{This figure describes the overall architecture of the UI Generator. The UI generator consists of a Transformer Encoder and a Transformer Decoder. The UI Generator first converts a text description into BERT Vectors and they get feeded with individual arrows into a block of Transformer Encoder at the top. Then, the UI Generator decodes new UI elements by taking all exsiting element's dimensions and coordinates attributes, and the class of the element and predict a mixture of gaussians distribution of next elements. This distribution is used to train the Transformer when compared with ground-truth next UI elements. The encoder and the decoder communicates through a cross attention mechanism.}
  \label{fig:arch}
\end{figure}

Once we have pairs of clean and sorted UI element attributes and classes, and text descriptions, we formulate the machine learning task as follows. Our model takes $k$ BERT vectors of a text description $t_{1...k}$ and generate $n$ UI elements $u_{1...n}$. Each UI element is represented by combining normalized X and Y coordinates and width and height values $[x, y, w, h]$ relative to the screen dimensions, with a one-hot vector $e^{(c)}$ that is 1 at the index of the semantic class number that the UI element belongs to, and 0 anywhere else. This gives us one $u_i = [x_i, y_i, w_i, h_i,e^{(c)}]$ vector for each UI element.

We can then consider this problem as a typical sequence-to-sequence translation problem where we `translate' a sequence of text tokens to a sequence of UI elements. We adopted the full original Transformer Encoder-Decoder \cite{transformer} architecture that has shown to be successful for this learning problem. As shown in Figure \ref{fig:arch}, our model architecture contains an encoder and a decoder sub-network that communicates through a cross-attention mechanism. The encoder only considers the text description  $t_{1...k}$ and consumes them solely as inputs. The decoder is an autoregressive model that takes all the previous UI elements $u_{1...i - 1}$ and the encodings from the encoder, and outputs the next UI element $u_i$ in the sequence at any particular time-step.

While we can train the decoder to generate exact coordinates and classes of UI elements, a generative model commonly outputs a distribution instead of a specific token in the inputs' format for each time-step. Because each UI element has continuous attributes $x, y, w, h$ and a discrete attribute $e^{(c)}$, we split the output of the decoder so that one part of it is used to parameterize Gaussian Mixture Models (GMM) for each continuous attribute, forming a Mixture Density Network \cite{mdn}. The other part of the outputs are treated as logits of a categorical distribution of the UI element class. At a high-level, this allows us to generate a distribution of output UIs instead of a single prediction. This is especially important for our problem because the text descriptions are often underspecified--a single text description, such as 'login page', can correspond to many potential candidate UI designs. Combining the encoder and decoder, the Transformer generates a probability distribution of each predicted output UI element:
$$ p(u_i) = p(u_i | \theta), u_i = [x_i, y_i, w_i, h_i,e^{(c)}], \theta = \textbf{Transformer}(u_{1...i-1}, t_{1...k})$$
To train this model, we minimize the total negative log likelihood of the ground-truth UI element sequence given the probability distribution above (note that we don't have to train $p(u_1)$ as it is always the start token).

$$ L(\textbf{Transformer}) = - \sum_{i = 2}^{n}\log(p(u_i)) $$

This is a typical training process for generating continuous, low-dimensional attributes with a sequence model, and we redirect interested readers to detailed descriptions of similar training processes in \cite{sketchrnn}. 

Both encoder and decoder in our final model have a hidden-layer size of 64, an intermediate size of 256, and 6 layers (chosen by performance on the validation set). We use an Adam optimizer with a starting learning rate of $10^{-3}$ and we manually decreased it to $10^{-4}$ and then $10^{-5}$ when the validation loss plateaued at each rate. We implemented our model using Trax and trained it on Google Cloud TPU v3 with 32 cores for ~7 days. 

\subsubsection{Sampling UIs}
\label{sec:sampling}
Once the model is trained, we can generate all the elements on a UI given a text description, in an auto-regressive manner. At each time-step we feed the Transformer all previously generated elements along with the text description, and sample the Transformer for the next predicted UI element $\hat{u_i}$ from the distribution $p(u_{i} | \textbf{Transformer}(u_{1...i-1}, t_{1...k}))$. We repeat the process until the model outputs a special token of \texttt{EOS} to indicate the end of generation. 
We added a temperature parameter of $0.1$ to both the categorical distribution of UI element classes (by rescaling the logits) and the GMM distribution (by rescaling the variances), which controls the stochasticity of the model that translates to diversity of generated UIs for a given text description. 


\subsubsection{Well-formedness Metrics, Filtering and Reranking}
\label{sec:filtering}
We adapted three existing well-formedness metrics defined in literature that are used as evaluation metrics to filter out low-quality UIs since they can appear by chance in our generative sampling approach. These metrics are respectively:

\begin{itemize}
    \item \emph{Overlap}--the percentage of area on the entire UI that is occupied by at least two elements.
    \item \emph{IOU}--the average intersection-over-union between any two elements in the UI.
    \item \emph{Alignment}--a metric introduced in \cite{neuraldesign} that approximates the average inter-element alignment distance between elements in the UI.
\end{itemize}

We computed the value for each of these metrics for each example in the validation set and filter out the generated UIs that exceed the average value in the validation set for any of the metrics. We then rerank the generated UIs by computing an overall score from the same three metrics. We first estimate each metric as separate probability distributions by modeling their values for all UIs in the validation set. We then find the candidate UI's score by taking their cumulative density function in the respective distributions, and multiplying these values together (this assumes each of the distributions are independent for each UI example). Finally, we take the top-50\%-scored UIs generated for each text description and use those as candidates of final outputs to users.


\subsubsection{Snapping-to-grid}
The last step of the generation process is to adjust the numeric outputs from the model to better aligned values. We snap each UI element to one of $32$ discrete grids along each dimension to remove minor alignment discrepancies as the model estimates continuous values for each of the continuous variables $x, y, w, h$. 

\subsection{Rendering}
The final step of all of our proposed methods is to render the produced UIs in the representation of low-fidelity mock-ups. Low-fidelity mock-ups are commonly used in the early stage of UI design, because they enable designers to focus on essential design ideas instead of being distracted by details.
We curated a set of SVG elements from the Material Design guidelines \cite{material}, and programmatically modified specific graphic attributes within these SVG elements for each individual UI element (e.g., only resizing the length of the bar of a slider instead of stretching the circle selector). These enhancements allow us to accommodate different aspect ratios of the UI elements to avoid stretching, producing coherent final UI mock-ups compiled in HTML documents. Note that for retrieval-based methods, we can optionally provide the original screenshots of the UIs in addition to the mock-ups as these UIs are taken directly from the dataset.
\section{Experiments}
To evaluate the effectiveness of each of our proposed methods in a standardized manner, we compare and contrast quantitative metrics on the screen2words test set across our methods. We also qualitatively evaluate UI examples suggested by each of our proposed methods based on individual text descriptions in the test set to gain high-level, qualitative insights into the characteristics of each method's results. Importantly, we realize each of the methods have their unique strengths and weaknesses in suggesting UIs, making them suited for different applications for different scenarios.

\subsection{Metrics}
\label{sec:metrics}
As described in Section~\ref{sec:filtering}, prior work has established a few metrics to evaluate the quality (well-formedness) of UIs. These metrics include Overlap, IOU, and Alignment. 


In addition to measuring the well-formedness of the generated/retrieved UIs, we believe a robust model also needs to preferrably suggest diverse and relevant UIs. While diversity measures introduced in prior work \cite{vtn} compare the overall distributions of the generated UIs and real UIs from the entire test set, our generated UIs are grounded on text descriptions, and hence cannot be evaluated using the same metrics. In addition, relevance (to given text descriptions) is an entirely new criterion for our proposed methods as prior generative models do not have the capability to generate UIs based on text descriptions.


We define both diversity and relevance metrics based on DocSim \cite{docsim}, a similarity measure designed for layouts and UIs. We consider the diversity of a set of UIs as the average pairwise DocSim measure between individual UIs in the set (this value decreases when diversity increases). To measure relevance, we take the maximum out of all DocSim metric values between each individual UI in the set of suggestions and the ground-truth paired UI of a particular text description.


\subsection{Quantitative Results}
\label{sec:quant}
\begin{table}
  \caption{Well-formedness, Diversity, and Relevance Metrics (*these are metrics reported by VTN \cite{vtn} on a different task with different data-processing methods, and should only be used for reference.)}
  \label{tab:freq}
  \begin{tabular}{l|ccc|c|c}
    \toprule
    &IoU $\downarrow$&Overlap $\downarrow$&Alignment $\downarrow$&Diversity $\downarrow$&Relevance $\uparrow$\\
    \midrule
    UI Generator & 0.115 & 0.294 & 0.600 & 0.0393 & 0.0757 \\
    Multi-modal Retriever & 0.0525 & 0.229 & 0.511 & 0.0309 & 0.0738\\
    Text-only Retriever & 0.0492 & 0.228 & 0.507 & 0.0167 & 0.0644\\
    Data (test set) & 0.0550 & 0.266 & 0.502 & - & - \\
    \midrule 
    (VTN \cite{vtn})* & (0.115) & (0.165) & (0.373) & - & -\\
  \bottomrule
\end{tabular}
\end{table}

To obtain an objective measure of each method's performance, we computed the aforementioned quantitative metrics on the test set for each of our proposed methods. We believe these metrics can be useful for future research to compare against our methods.

For well-formedness metrics of both retrieval methods, we included these metrics primarily for reference purposes, as these metrics are computed from real UIs in the training set and therefore they are highly similar to the average metric of the training set.

For well-formedness metrics of the UI Generator, we sample 10 UIs for each text description in the test set, and compute the average metrics over all of the sampled UIs. We observed that while the UI Generator is not able to outperform ground-truth data and retrieval-based methods on the metrics, it was able to generate UIs that approach the other conditions' UIs' well-formedness. As a reference, we also included the results reported by VTN \cite{vtn}, a state-of-the-art prior work in layout generation, but note that those results are not directly comparable with ours because their task does not involve text descriptions and they process the datasets differently. 

For diversity and relevance measures, the UI Generator generates less diverse but more relevant sets of UIs compared to the other methods. It is possible that retrieval-based methods make forced decisions for picking top-N matches, which might return more diverse UIs that might be less relevant to the given text description (see the high Diversity and low Relevance of the Text-only Retriever in Table~\ref{tab:freq}). In contrast, the UI Generator can more smoothly interpolate between different types of UIs that relate to the same text description, hence generating UIs that have higher chances of being relevant to the ground-truth. We further observe these trends in sets of results investigated in Section~\ref{sec:qual}.

\subsection{Qualitative and Comparative Analysis}
\label{sec:qual}
\begin{figure}[h]
  \centering
  \includegraphics[width=1.0\linewidth]{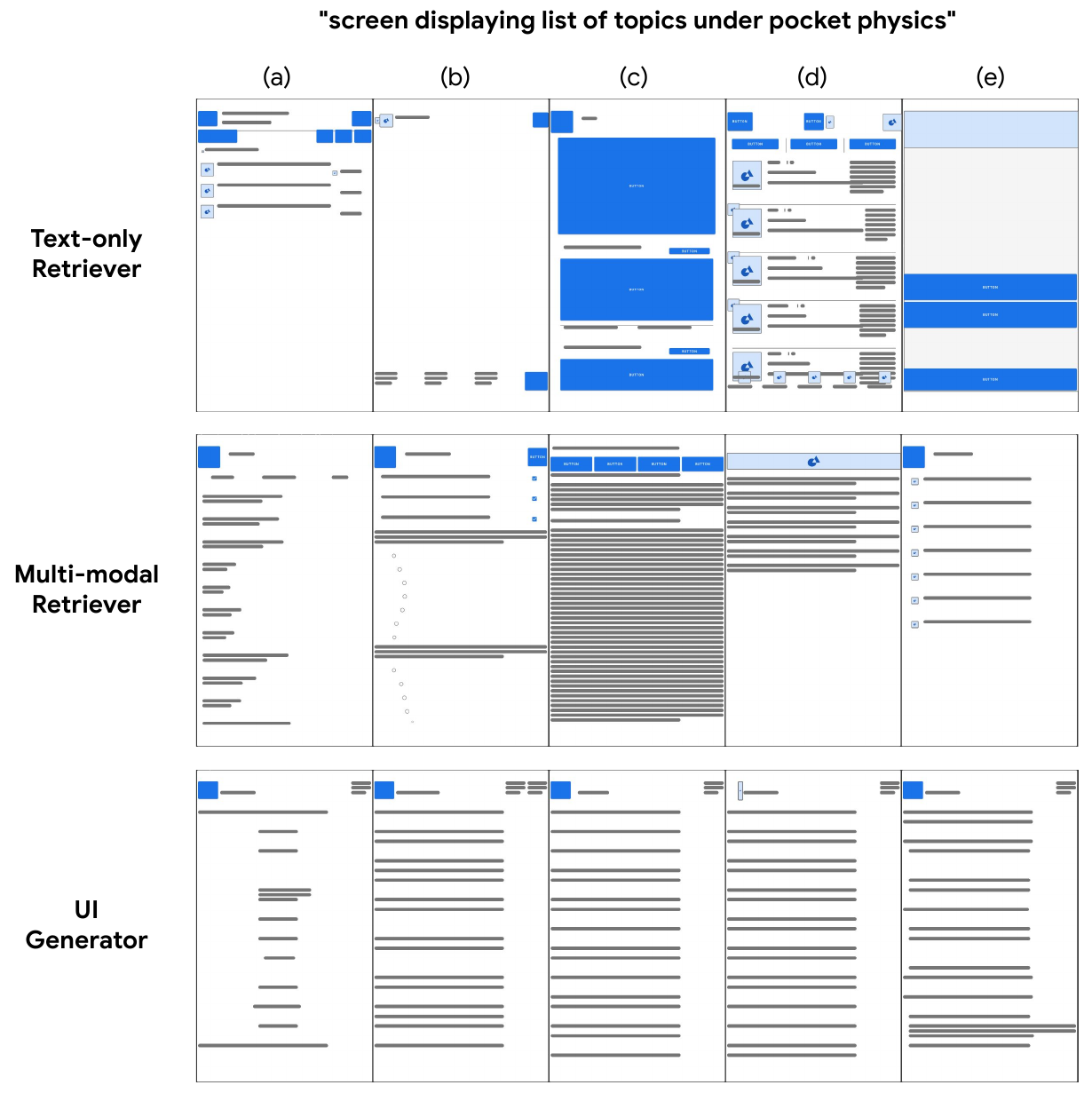}
  \caption{UI mock-ups created by our methods in response to the text description \textit{``screen displaying list of topics under pocket physics''}.}
  \Description{This figure shows a three by five grid of UIs generated by our proposed methods given the description of screen displaying list of topics under pocket physics. The top row are UIs generated by the text-only retriever, which has the most diversity. The middle row are UIs generated by the Multi-modal Retriever, which best corresponds to the description, The bottom row are generated by the UI Generator which are variations of similar lists.}
  \label{fig:list}
\end{figure}

\begin{figure}[h]
  \centering
  \includegraphics[width=1.0\linewidth]{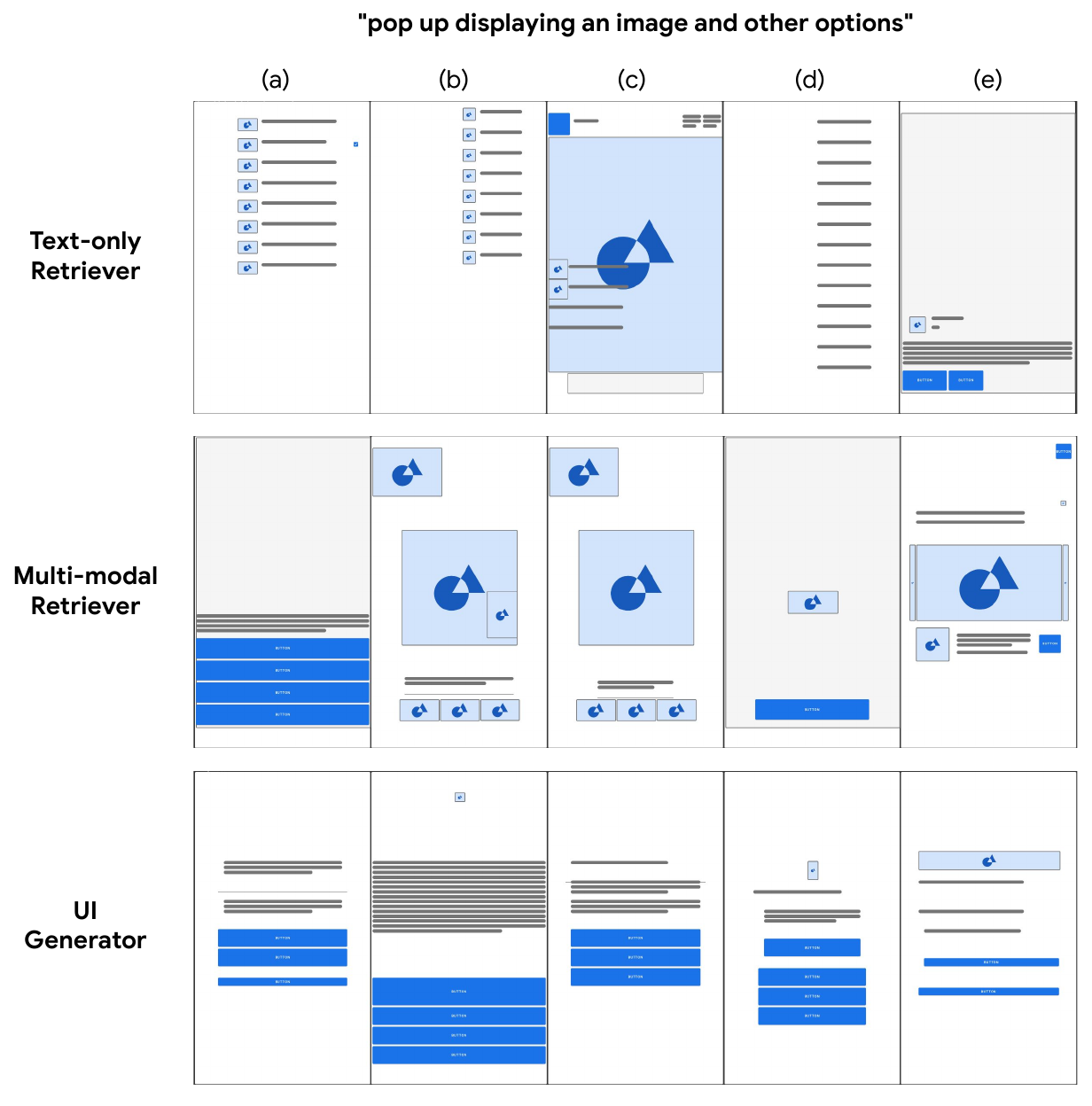}
  \caption{UI mock-ups created by our methods in response to the text description \textit{``pop up displaying an image and other options''}.}
  \Description{This figure shows a three by five grid of UIs generated by our proposed methods given the description of pop up displaying an image and other options. The top row are UIs generated by the text-only retriever, which did not retrieve relevant pop up interactive UIs with images but retrieved pop up menus. The middle row are UIs generated by the Multi-modal Retriever, which best corresponds to the and has multiple UIs with a large image, The bottom row are generated by the UI Generator which are variations of similar pop up screens but most look plausible as pop-up UIs.}
  \label{fig:popup}
\end{figure}

Beyond obtaining quantitative metrics for each set of methods for concrete reference statistics, we qualitatively investigated the characteristics of results from our proposed methods. Overall, we observe that most UI mock-ups created by our methods are well-formed and comprehensible, and are within the general categories of UIs that we would expect based on the text descriptions. In this section, we analyze the results acquired by each method for two text descriptions from the test set\footnote{These captions are also within the set that was used in Section 5, which were chosen with mock-up results unseen to us prior to selection.}: \textit{``screen displaying list of topics under pocket physics''} (see Figure~\ref{fig:list}) and \textit{``pop up displaying an image and other options''} (see Figure~\ref{fig:popup}). 

\subsubsection{Text-only Retriever Results}
We observed that Text-only Retriever is able to make large conceptual shifts across various examples. From the results in Figure~\ref{fig:list}, we see very diverse types of lists, e.g., a list with large buttons (c), a list with thumbnails (d), and a list that is only partially filled (a). These results represent a very diverse set of designs, and we believe this is because the model reasons merely at the abstract level of natural language descriptions. However, such high diversity also leads to the retrieval of some irrelevant UI examples, such as the UI with an empty screen (Figure \ref{fig:list}b) and the UI with only a few buttons (Figure \ref{fig:list}e). Moreover, because the Text-only Retriever was not trained specifically on UI captions, it might not be able to recognize more detailed, domain-specific specifications of UIs. For instance, in Figure~\ref{fig:popup} the Text-only Retriever was only able to gather a set of drop-down/pop-up menus, while a ``pop up displaying an image'', as seen from other examples, typically represents a popup window with interactable elements instead of a menu.

\subsubsection{Multi-modal Retriever Results}
The Multi-modal retriever, in contrast to the Text-only Retriever, excels at matching specific details described in text. For instance, only this method was able to retrieve pop-ups with emphasized large images specified in \textit{``pop up displaying an image...''}. We believe this is because the Multi-modal Retriever has access to text-based features of individual elements and text descriptions in the UIs. Nevertheless, this method also gets access to content-specific text details, which might be confusing for retrieval given some text descriptions. For example, the model retrieved a UI with a large block of text that seemingly conflicts with \textit{``a list of topics''} at Figure~\ref{fig:list}c. However, this UI was originated from an app presenting business topics and ideas which contain text content related to ``lists''. This provides an example where matching text-content on the screen and within the app might not necessarily lead to matching design concepts.

\subsubsection{UI Generator Results}
For the results synthesized from scratch by the UI Generator, we observe that this method is good at making small but coherent variations within the same theme for both cases in Figure~\ref{fig:list} and ~\ref{fig:popup}. Figure~\ref{fig:list} shows minor changes the UI Generator made to vary list structures. Similarly, we observe popups with varying layouts that follow the same theme in Figure~\ref{fig:popup}. Nevertheless, such subtle variations can lead to the lack of large thematic variations and lower diversity of UIs in these sets of results. The lack of variations is also observed across the entire test set as shown by the diversity metric in Section~\ref{sec:quant}. However, this issue can be addressed by hyper-parameter tuning: the UI Generator contains a tunable sampling temperature hyper-parameter (currently set to a fixed value of $0.1$) that can be explored in the future to vary the balance between diversity and well-formedness of the UIs. It is worth noting that while it is easy for retrieval-based methods to return well-formed UIs, it is nontrivial for a generative method to return sensible UIs because such method needs to synthesize complete UIs from scratch, instead of using an existing UI.


\section{User Study}
\label{sec:study}
To further solidify our assessments on the characteristics of results gathered by each of our proposed methods in Section~\ref{sec:qual}, we conducted a user study to gauge professional UI designers and practitioners' preferences on mock-ups created by our methods.

\subsection{Procedure}

To solicit our participants' preferences for the mock-ups our methods have created, we manually chose five text descriptions from the test dataset that cover diverse design scenarios. We then used each of our proposed methods to create 5 UI mock-ups for each of these descriptions. We \emph{did not inspect} the mock-ups created by these methods prior to selecting these descriptions. 

This allows us to form 3 sets of 5 UI mock-ups for each of the 5 text descriptions. In the study, we frame each text description as a design goal. The participants were first presented with a ground-truth high-fidelity screenshot corresponding to the design goal in text. They were then presented with a scrambled 3x5 grid of all 15 UI designs generated across three methods. The position of these design mock-ups are randomized. 
Based on these UIs, they would then answer the following two questions and explain their rationales.

\begin{itemize}
    \item Q1. Which mock-up in the collection best resembles the hi-fidelity design?
    \item Q2. Select the mock-up in the collection that presents the best design alternative for the given design goal.
\end{itemize}

At the end of the survey, we also obtained the participants' opinions on the following questions on AI-assisted design in general, which is finally followed by them providing open-ended comments or suggestions for this work.
\begin{itemize}
    \item Qa. If an AI system is built to suggest UIs from text description inputs just as the ones presented in this study, how might it help / not help with your design workflow?
    \item Qb. Following the above question, if such an AI system that suggests UIs from text is developed, what do you think are the most important features this system should have?
\end{itemize}

\subsection{Participants}
We recruited 15 professional UI/UX practitioners to complete this survey from a mailing-list internal to our organization. The participants self-reported an average of ~9.8 years of industrial UI/UX design experience. Their professions include UI/UX Designer, UX Engineer, Interaction Designer, UX Researcher, and are located in the United States and the United Kingdom. All participants completed the survey at their own pace remotely, and were each compensated with a \$20 USD gift card upon completion of the survey. Each participant took roughly 20 minutes to complete the survey.

\subsection{Results \& Discussion}
We split our analysis of the study results by the five design goals. Q1s of each design goal serve as warm-up questions for participants to familiarize themselves with the mock-ups' representation. These questions also gauge the created UIs' relevance and diversity as perceived by designers, because a relevant and diverse set of UI suggestions has a higher chance of covering the high-fidelity UI associated with the design goal. We found that for a given design goal (description), each of our methods is more frequently chosen by participants than other methods (Chi-square goodness-of-fit test, $p < 0.05$ for each case, with equal expected frequency across methods) once, while results are insignificant for the other two descriptions. This result shows that each of the methods is competitive for specific scenarios.

\begin{figure}[h]
  \centering
  \includegraphics[width=1.0\linewidth]{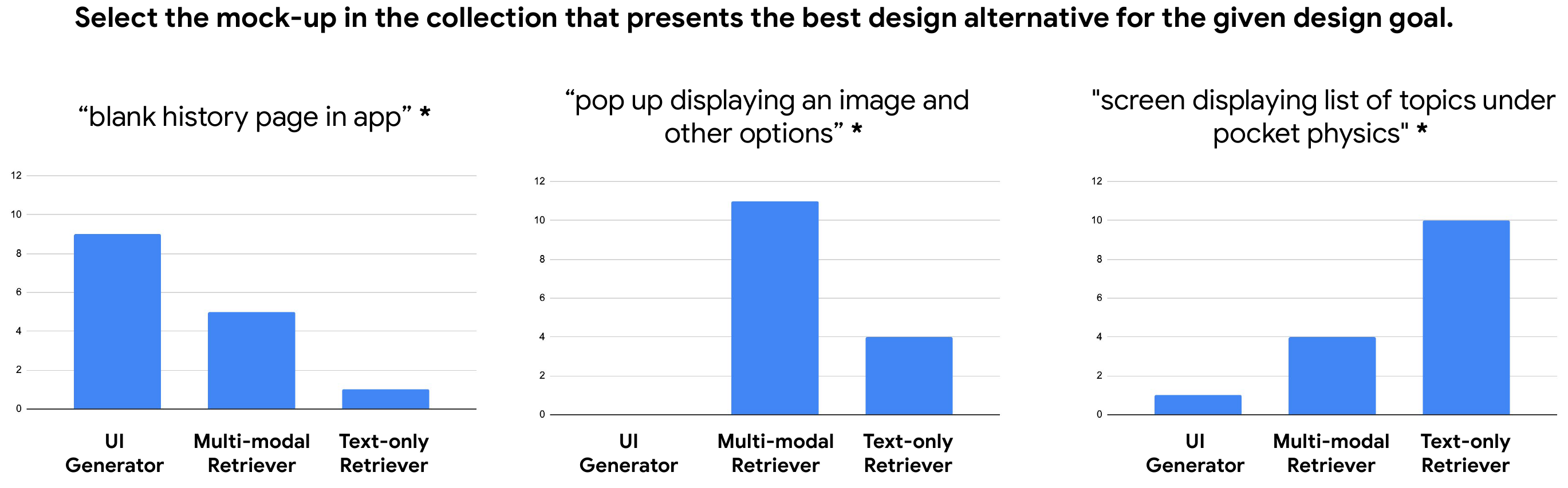}
  \caption{The study results for best alternative designs for three of the five design goals (Q2.). We observe that each of our  methods is preferred in one of the design goals. The design goals are shown at the top of each chart.}
  \Description{This figure shows a three bar charts that compares the frequency that each of the methods gets voted as the best UIs. For blank history page in app on the left, the UI generator gets voted 9 times, as opposed to 5 and 1 time for multi-modal retriever and text-only retriever. For pop up displaying an image and other options in the middle, the UI generator gets voted 0 times, as opposed to 11 and 4 times for multi-modal retriever and text-only retriever. For screen displaying list of topics under pocket physics on the right, the UI generator gets voted 9 times, as opposed to 5 and 1 time for multi-modal retriever and text-only retriever. For pop up displaying an image and other options in the middle, the UI generator gets voted 1 times, as opposed to 4 and 10 times for multi-modal retriever and text-only retriever. All results are significant according to the Chi-square goodness of fit test.}
  \label{fig:study}
\end{figure}

Q2s in our study are more important for evaluating the overall quality and relevance of UI suggestions. Similar to the first question, each method was more frequently chosen by participants ($p < 0.05$) than the other methods in one of the three descriptions, while the other two descriptions have insignificant results. The particular designs chosen by participants align with our understanding of the characteristics of each of the methods presented in Section~\ref{sec:qual}. For instance, our participants considered the Multi-modal Retriever to have provided the best design alternatives (Figure~\ref{fig:popup}) for \textit{``pop up displaying an image and other options''} (the middle chart in Figure~\ref{fig:study}). They prefer these mock-ups due to their high relevance to the text description for including an image, as commented by a participant (P8): \textit{``[this design] allows users to focus on the main image''}. 

In another example, the results based on the description \textit{``blank history page in app''} (see Figure~\ref{fig:study}) indicate that most participants preferred the mock-ups created by the UI Generator, citing that it is nice to have an empty state illustration instead of nothing (P4). The mock-up generated by the UI Generator still retains the general structure of an empty UI but with an extra large illustration. This result showed the benefit of the generative method which allows variations under the same theme given a text description.  

While we have evaluated a diverse set of UI suggestions and design goals, we acknowledge that our evaluation is limited in that we only inspected a small set of design cases that can potentially benefit from our methods. We suggest future work to more comprehensively understand general trends that exist across large-scale datasets for these methods. The open-ended questions regarding AI-assisted design systems at the end of the survey also provided us with a number of actionable suggestions for developing our methods into AI-driven systems. We discuss participants' responses to these questions in the following section for potential applications (Section~\ref{sec:applications}). 

\section{Potential Applications}
\label{sec:applications}
Our ultimate goal of developing these methods is to provide intelligent design support for UI designers and developers with relevant design artifacts given high-level text descriptions. We discuss several potential applications that can utilize our methods based on open-ended comments provided by our participants in the study.

\subsection{Early-stage Design Sketch Rendering}
A major theme emerged from the participants' feedback for Qa. is that the UI suggestions presented to participants would be good for initial stages of design, such that they can be good starting points for higher-fidelity designs in their design processes. P2 mentioned that these designs can be \textit{``providing some quick, initial mocks that I could leverage for higher fidelity design solutions early on in my design process.''} P9 commented that \textit{``this can in turn help in brainstorming and generating a greater variety of designs.''} As such, one potential application is to further develop systems catered specifically for early-stage design. For instance, we can adjust our rendering engine to display sketch-like mock-ups, which would encourage more high-level early feedback. Sketch-based rendering would allow designers to discuss these generated designs as early-stage artifacts and interact with them directly with pen strokes. As mock-ups created by our methods contain attributes of each UI element, this application can be realized by replacing the set of rendering elements with sketched elements introduced in the UISketch dataset \cite{uisketch}. 



\subsection{Interactive and Steerable UI Generation}
On the other hand, one of the most requested features by the participants, as reflected by their responses to Qb., is to give designers more control over these AI-assisted methods. P7 mentioned that it \textit{``should be able to generate other layouts and substitutions if I pin certain controls or elements.''} Similarly, P8 requested these methods to \textit{``allow people to quickly modify the AI-generated design and maybe even control the design directions.''} These can be achieved by further developing the UI Generator as it generates one new UI element at a time based on all previously generated elements. We can simply feed elements that are already designed by the designer, instead of only using the model's previously predicted elements, as inputs to the autoregressive decoder. This enables designers to control the design process with their own designs. Moreover, as the UI Generator outputs UI elements using a sampling-based approach, it can generate multiple candidate elements each time and allow designers to select their preferred element. Designers can also adjust the candidate diversity of the model by changing the temperature parameter that controls the stochasticity of the sampling process as discussed in Section~\ref{sec:sampling}. These features could provide designers with rich control over the mock-up creation process as requested in the user study. 



\subsection{Combining Multiple UI Suggestion Methods}
As we discovered in the previous sections, each of our methods is suitable for obtaining designs of varying levels and types of diversity and hence are suited for various stages of the design process, we envision a final system that combines all of these methods to support different stages of the design process. A designer might start with the Text-only Retriever for high-level design inspirations from the mock-ups and optionally concrete UI screenshots examples that it can retrieve. With the gradual convergence of design details, the system can use the Multi-modal Retriever to retrieve more closely related and specific designs. Finally, the designer can use the UI Generator to innovate on design artifacts based on existing designs they prefer. 

\section{Discussion \& Future Work}
We discuss several design choices and limitations of our proposed methods in this section. These aspects of the current version of our proposed methods lead to several avenues of future work that we believe researchers could pursue.

\subsection{Modeling Choices}
There are a number of architecture choices that can be further explored for our proposed methods. For example, for methods that encode text and UIs separately, we only used the same number of layers and hidden units in each layer for the two sub-networks that encode text and UIs respectively (Multi-modal Retriever and UI Generator). However, we believe each of these sub-networks should be tuned independently for an optimal configuration, especially when modeling text descriptions might require different levels of abstraction from modeling UI elements. While all our methods use separate sub-networks to encode data of each of the modalities, more recent approaches have used a single decoder-only architecture (e.g., DALL-E \cite{dalle}) even for multi-modal generation tasks. Using a decoder-only architecture might encourage the model to more effectively form cross-modality correspondences that could be helpful for suggesting UIs based on more low-level, structured and detailed text descriptions.


\subsection{Supporting Design-Specific Language}
Another limitation of our work is related to partial misalignment between the descriptions in the dataset that we use and the type of descriptions we aim to support. The screen2words dataset is collected by asking crowd-workers to summarize UI screenshots in natural language. The type of language that crowd-workers use to describe UI screenshots might not be perfectly aligned with how designers would articulate design goals. For example, a text description in this dataset might include content-based details which could be less important to the design goal, or lack information about design constraints that a designer might desire.
Nevertheless, our work made a valuable contribution by investigating three novel methods, which are applicable to other paired text-UI datasets that might become available in the future. Investigating the degree of such misalignment and the effectiveness of our methods on design-specific language remain future work.



\section{Conclusion}
This paper presented three deep-learning methods for creating low-fidelity UI mock-ups from high-level natural language descriptions, including one generative model and two retrieval models. 
To our knowledge, all of these methods are the first in their kinds to achieve the affordance of text-based UI design creation. We compare and contrast the results produced by each of these methods using quantitative metrics and qualitative investigations, and offer an analysis into the characteristics of each method. We further conducted a user study with 15 professional UI designers and practitioners, which confirmed our findings for the characteristics of these methods and revealed several important directions for future work. We believe these methods and our findings are important first steps towards language-based design generation to assist designers in their design practices.

\begin{acks}
We would like to thank all our recruited UI designers and practitioners for participating in our study. We would also like to thank Tao Dong, James Lin, David Ha, Eldon Schoop, Sarah Sterman, and Mingyuan Zhong for providing valuable feedback to the project.
\end{acks}

\bibliographystyle{ACM-Reference-Format}
\bibliography{acmart}

\end{document}